\pdfoutput=1

\documentclass[11pt]{article}

\usepackage{acl}

\usepackage{times}
\usepackage{latexsym}

\usepackage[T1]{fontenc}

\usepackage[utf8]{inputenc}

\usepackage{microtype}

\usepackage{inconsolata}

\usepackage[skins]{tcolorbox}

\usepackage{amsfonts}
\usepackage{amsmath}
\usepackage{multirow,makecell}
\usepackage{booktabs}
\usepackage{graphicx,subfigure}
\newcommand{\hide}[1]{} 
\def\code#1{\texttt{#1}}

\def\bd{{\mathbf{d}}}

\def\RR{{\mathbb{R}}}

\title{Beyond Yes and No: Improving Zero-Shot LLM Rankers via Scoring Fine-Grained Relevance Labels}

\author{
    Honglei Zhuang, Zhen Qin, Kai Hui, Junru Wu, Le Yan, \\
    \bf{Xuanhui Wang \and Michael Bendersky} \\
    Google Research \\
    \texttt{\{hlz,zhenqin,kaihuibj,junru,lyyanle,}\\
    \texttt{xuanhui,bemike\}@google.com}     
}

\begin{document}

\maketitle
\begin{abstract}

Zero-shot text rankers powered by recent LLMs achieve remarkable ranking performance by simply prompting.
Existing prompts for pointwise LLM rankers mostly ask the model to choose from binary relevance labels like ``Yes'' and ``No''. 
However, the lack of intermediate relevance label options may cause the LLM to provide noisy or biased answers for documents that are partially relevant to the query.
We propose to incorporate fine-grained relevance labels into the prompt for LLM rankers, enabling them to better differentiate among documents with different levels of relevance to the query and thus derive a more accurate ranking.
We study two variants of the prompt template, coupled with different numbers of relevance levels.
Our experiments on 8 BEIR data sets show that adding fine-grained relevance labels significantly improves the performance of LLM rankers.


\end{abstract} 

\section{Introduction}
\label{sec:intro}

Large language models (LLMs) such as GPT-4~\cite{gpt4} and PaLM 2~\cite{palm2} have demonstrated impressive zero-shot performance on a variety of NLP tasks.
Recently, there has been a growing interest in applying LLMs to zero-shot text ranking, with remarkably impressive results.
The earliest zero-shot LLM rankers are pointwise~\cite{liang2022holistic,sachan2022improving}, which score one query and one document at each time and rank the documents based on the scores.
Lately, pairwise~\cite{qin2023prp} and listwise~\cite{sun2023chatgpt,ma2023listwise} LLM rankers also show strong performance, but they cannot scale to long lists and still largely rely on a high-quality first-stage ranking.

A typical category of pointwise LLM rankers is relevance generation~\cite{liang2022holistic}. 
In this method, the LLM is prompted to answer whether a document is relevant to the query.
Existing pointwise LLM rankers mostly ask the LLM to answer ``Yes'' or ``No'' and use their likelihood to derive a ranking score.
Nevertheless, some documents cannot always be accurately classified into these two categories as they may not directly answer the query but still contain helpful information.

Studies on human subjects show that using binary options sometimes leads to biased answers~\cite{rivera2022continuous}.
Instead, providing reasonably fine-grained options can lead to more reliable results~\cite{roitero2018fine_grained,birkett1986selecting,rivera2022continuous,johnston2017contemporary}.
Actually, in information retrieval data sets, the annotation guidelines for human annotators often employ multiple relevance levels, like the 3-level scale used in TREC-COVID~\cite{voorhees2021trec} and TREC-Robust~\cite{robust04}, as well as the 4-level scale used in TREC-DL~\cite{trecdl19,trecdl2020}.
We believe that a zero-shot LLM ranker might share the same behavior pattern with human annotators.

Therefore, we propose to explicitly provide fine-grained relevance labels in the prompt to zero-shot LLM rankers.
Instead of asking the LLM to choose between two options, we provide the LLM with fine-grained relevance labels, such as ``Highly Relevant'', ``Somewhat Relevant'' and ``Not Relevant'' and collect their likelihood scores from LLM predictions to derive the ranking score.
The intuition is that the intermediate relevance labels in the prompt serve as a ``cue'' to the LLM to distinguish partially relevant documents from fully relevant or fully irrelevant ones.

Our evaluation on 8 BEIR~\cite{thakur2021beir} datasets demonstrates that simply adding intermediate relevance labels significantly boosts LLM ranking performance across different datasets, regardless of the actual ground-truth label granularity. 
An in-depth analysis reveals that the proposed new prompt enables LLM rankers to distinguish documents previously indistinguishable with the binary-option prompt.

\section{Related Work}
\label{sec:related}

\paragraph{Zero-shot LLM rankers.}
Shifted from tuning-based learning to rank on textual and traditional tabular datasets~\cite{monobert, tfrbert, zhuang2021ensemble, nogueira2020T5ranking, zhuang2023rankt5, xian2022learning, liu2009ltr, qin2021AreNeuralRankers}, 
there is an emerging thread of research exploring how to use general-purpose LLMs directly or indirectly~\cite{jagerman2023query,li2023generate} for zero-shot text ranking.

\citet{liang2022holistic} and \citet{sachan2022improving} adopt a pointwise approach which scores the relevance of one document at a time based on how likely the LLM would classify the document as relevant or how likely the LLM would generate the query from the document respectively.
There are also explorations on pairwise~\cite{qin2023prp} and listwise~\cite{sun2023chatgpt,ma2023listwise,zhuang2023setwise} LLM rankers which take multiple documents as input and return the ranking directly, but they are usually applied iteratively on smaller sets of documents.
In this paper, we only focus on pointwise LLM rankers.

\paragraph{Zero-shot LLM assessors.}
Another related research area~\cite{faggioli2023perspectives,thomas2023large} employs LLMs as assessors, where fine-grained relevance labels are also provided in the prompt.
However, these methods do not use the likelihood scores of fine-grained relevance labels.
The goal of LLM assessors is to provide a relevance label for every query-document pairs that aligns with the ground-truth relevance label, potentially created by human assessors.
LLM assessors are usually used to create an evaluation data set, which can be used to reliably evaluate different ranking models.
This is different from LLM rankers, which typically only need to ensure that the relative order of the top-ranked documents are accurate.

\begin{figure}[!t]
  \centering
  \subfigure[Yes-No relevance generation]{
    \label{subfig:intro_fig_1}
    \includegraphics[width=1.0\columnwidth]{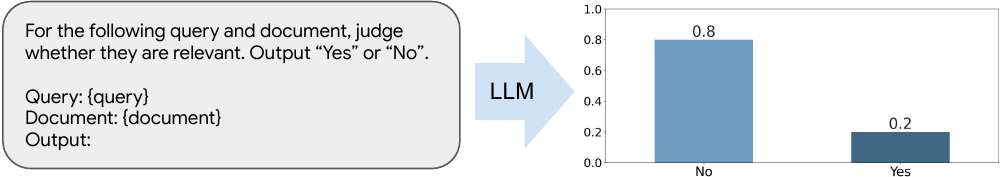}
  }
  \subfigure[Fine-grained relevance label generation]{
    \label{subfig:intro_fig_2}
    \includegraphics[width=1.0\columnwidth]{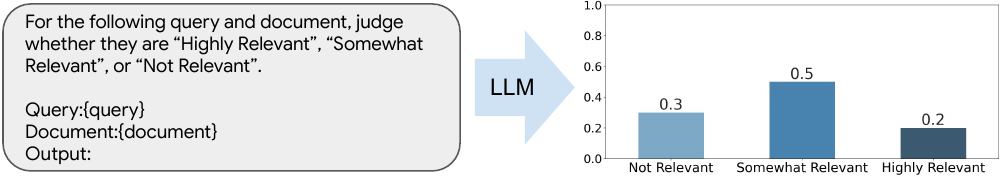}
  }
  \subfigure[Rating scale relevance generation]{
    \label{subfig:intro_fig_3}
    \includegraphics[width=1.0\columnwidth]{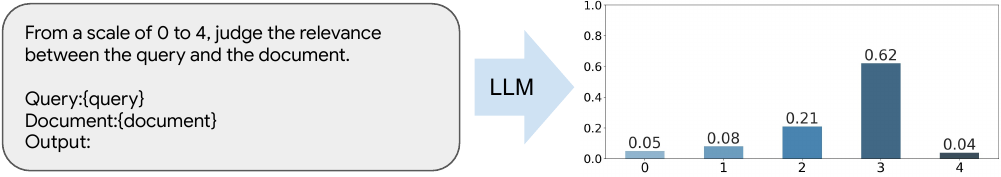}
  }
  \caption{
  Illustration of different prompting strategies for relevance generation LLM rankers.
  }
  \label{fig:intro}
\end{figure}

\section{LLM Rankers}
\label{sec:model}

\subsection{Preliminaries}

Existing explorations using zero-shot LLMs as pointwise rankers can be broadly divided into two categories: relevance generation~\cite{liang2022holistic} and query generation~\cite{sachan2022improving}. We focus on relevance generation in this work.

Given a query $q$ and a list of candidate documents $\bd = (d_{1}, \ldots, d_{m})$,
an LLM ranker based on relevance generation takes each query-document pair $(q, d_i)$ as input and prompts the LLM to answer whether the document is relevant to the query by ``Yes'' or ``No'' (see Figure~\ref{subfig:intro_fig_1}). 
Then a ranking score $f(q, d_i) \in \RR$ for each document is calculated based on LLM's log-likelihood score $s_{i,1} = \text{LLM}(\text{Yes}|q, d_i)$ and $s_{i,0} = \text{LLM}(\text{No}|q, d_i)$ by using a softmax function~\cite{nogueira2020T5ranking}:
\begin{align}
    f(q, d_i) = \frac{\exp(s_{i,1})}{\exp(s_{i,1}) + \exp(s_{i,0})} \nonumber
\end{align}
The ranked list is obtained by sorting the documents based on their ranking scores.

\subsection{Prompts}
In many datasets, there exist documents that are only partially or marginally relevant to the query, which LLMs struggle to classify into two classes.

\paragraph{Fine-grained relevance labels.}
We extend the classical relevance generation methods by introducing fine-grained relevance labels.
Without loss of generality, we use a set of 3-level graded relevance labels as example: [``Not Relevant'', ``Somewhat Relevant'', ``Highly Relevant''], denoted as $[l_0, l_1, l_2]$.
Then, for each query-document pair $(q, d_i)$, we ask the LLM to evaluate their relevance by choosing from the given relevance labels.
We can obtain the log-likelihood of the LLM generating each relevance label:
\begin{align}
    \label{eq:llm_loglikelihood}
    s_{i,k} = \text{LLM}(l_{k}|q, d_i)
\end{align}
This example is illustrated in Figure~\ref{subfig:intro_fig_2}.
The exact prompt can be found in Appendix~\ref{sec:appendix_prompts}.

\paragraph{Rating scale.}
To avoid using relevance labels with potentially ambiguous order, we can also employ a rating scale. For example, we can prompt the LLM to rate the relevance between the query $q$ and the document $d_i$ on a scale from 0 to 4.
We can then use the LLM to obtain the log-likelihood $[s_{i,0}, \ldots, s_{i,4}]$ of generating each relevance scale value $[l_0, \ldots, l_4]$, which are ``0'' to ``4'' respectively.
This method allows us to try arbitrarily fine-grained relevance levels in the prompt.
Figure~\ref{subfig:intro_fig_3} illustrates an example of this prompt.
The exact prompt can be found in Appendix~\ref{sec:appendix_prompts}.

\subsection{Ranking Scores}
\label{sec:ranking_score_derivation}
Once we obtain the log-likelihood of each relevance label, we can derive the ranking scores.

\paragraph{Expected relevance values (ER).}
The most straightforward way is to calculate the expected relevance value.
First, we need to assign a series of relevance values $[y_0, y_1, y_2]$ to all the relevance labels $[l_0, l_1, l_2]$, where $y_k \in \RR$. 
Then we can calculate the expected relevance value by:
\begin{align}
    \label{eq:expected_relevance}
    f(q, d_i) &= \sum p_{i,k} \cdot y_k \\
    \text{where}~p_{i,k} &= \frac{\exp(s_{i,k})}{\sum_{k'} \exp(s_{i,k'})}  \nonumber
\end{align}

The relevance values $y_k$ can be provided by users or even tuned based on a training data set.
We empirically find that na\"{i}vely assigning $y_k = k$ (with $l_0$ to $l_k$ ordered from least to most relevant) already yields excellent performance.
Therefore, we simply adopt $y_k = k$.

\paragraph{Peak relevance likelihood (PR).}

We can further simplify ranking score derivation by focusing on top-ranked items.
We propose to only use the log-likelihood of the peak relevance label (``Highly Relevant'' in this example).
More formally, let $l_{k^*}$ denote the relevance label with the highest relevance.
We can simply rank the documents by:
\begin{align}
    \label{eq:max_relevance}
    f(q, d_i) = s_{i,k^*}
\end{align}
Note that $s_{i,k^*}$ is the log-likelihood directly obtained from the LLM, instead of the marginal probability $p_{i,k^*}$ in Equation~\eqref{eq:expected_relevance}.
Hence, it is not necessary to score all relevance labels using the LLM and could potentially save some decoding cost when using this strategy to derive the ranking score.
While this method is shown less effective on smaller models~\cite{nogueira2020T5ranking}, it works well empirically with larger models in our experiments.

\section{Experiment Setup}
\label{sec:exp}

\paragraph{Data set.}
We conduct experiments on 8 chosen data sets~\cite{sun2023chatgpt} from BEIR~\cite{thakur2021beir}: Covid, Touche, DBPedia, SciFact, Signal, News, Robust04, and NFCorpus.
Notice that our method is applicable regardless of the actual relevance granularity in each data set. 

We use BM25~\cite{Lin_etal_SIGIR2021_Pyserini} to retrieve the top-100 documents for each data set, and then rank the retrieved documents using LLMs with our proposed methods. 
We use FLAN PaLM2 S~\cite{palm2} as the LLM in our main experiments.
Results of other LLMs can be found in Appendix~\ref{sec:other_llms}.

The ranking performance is measured by NDCG@10~\cite{jarvelin2002cumulated}.

\begin{table}[!t]
    \centering
    \caption{
        Relevance labels used in RG-$k$L. The relevance label $l_{k^*}$ with the maximum relevance value is bolded.
    }
    \scalebox{0.8}{
        \begin{tabular}{c|p{0.9\columnwidth}}
        \toprule
         Method & Relevance Labels \\
        \midrule
        RG-2L & ``Not Relevant'', \textbf{``Relevant''} \\
        \midrule
        RG-3L & ``Not Relevant'', ``Somewhat Relevant'', \textbf{``Highly Relevant''} \\
        \midrule
        RG-4L & ``Not Relevant'', ``Somewhat Relevant'', ``Highly Relevant'', \textbf{``Perfectly Relevant''} \\
        \bottomrule
        \end{tabular}
    }
    \label{tab:labels used}
\end{table}

\paragraph{Compared methods.}

We compared the following prompting strategies:
\begin{enumerate}
    \item Query Generation (QG). Ranking documents based on the query likelihood from LLM given the document~\cite{sachan2022improving}.
    \item Binary Relevance Generation (RG-YN). Prompting the LLM with a query-document pair and using ``Yes/No'' likelihood to calculate the ranking score~\cite{liang2022holistic}. 
    \item $k$-Level Relevance Generation (RG-$k$L). Prompting the LLM to choose from $k$ relevance labels for each query-document pair. The relevance labels are listed in Table~\ref{tab:labels used}.
    \item Rating Scale $0$-to-$k$ Relevance Generation (RG-S$(0,k)$). Prompting the LLM to rate the relevance for each query-document pair using a scale from 0 to $k$. Note that for RG-S$(0,k)$, the LLM needs to score $(k+1)$ labels.
\end{enumerate}

By default, the ranking scores of our methods are derived using expected relevance (Equation~\eqref{eq:expected_relevance}).

\begin{table*}[ht!]
    \centering
    \caption{
        Overall ranking performances measured by NDCG@10 on BEIR data sets. 
        The best performances are bolded. 
        Average results that are significantly (paired $t$-test, $p<0.05$) better than RG-2L are marked with $^{*}$.
    }
    \scalebox{0.8}{
        \begin{tabular}{c|cccccccc|c}
        \toprule
         Method & Covid & Touche & DBPedia & SciFact & Signal & News & Robust04 & NFCorpus & Average \\
        \midrule
        QG & 0.7357	& 0.2408& 0.3773& 0.7495& 0.2872& 0.4156& 0.4651	&0.3673& 0.4548 \\
        RG-YN & 0.7897 & 0.2427 & 0.3696 & 0.6958 & 0.3196 & 0.4588 & 0.5656 & 0.3743 & 0.4770 \\
        \midrule
        RG-2L & 0.7949 & 0.2411 & 0.3590 & 0.7290 & 0.2996 & 0.4623 & 0.5636 & 0.3814 & 0.4789 \\
        RG-3L & \textbf{0.8065} & 0.2650 & 0.4013 & 0.7671 & 0.3142 & \textbf{0.4890} & 0.5660 & 0.3849 & 0.4992$^{*}$ \\
        RG-4L & 0.8063 & 0.2388 & 0.4033 & \textbf{0.7766} & 0.3184 & 0.4884 & 0.5635 & 0.3801 & 0.4969$^{*}$ \\
        \midrule
        RG-S$(0,2)$ & 0.7760 & 0.2695 & 0.3709 & 0.6921 & 0.3034 & 0.4677 & 0.5557 & 0.3787 & 0.4768 \\
        RG-S$(0,4)$ & 0.8048 & \textbf{0.2757} & \textbf{0.4190} & 0.7521 & \textbf{0.3301} & 0.4790 & \textbf{0.5668} & \textbf{0.3901} & \textbf{0.5022}$^{*}$ \\
        \bottomrule
        \end{tabular}
    }
    \label{tab:overall_comparison}
\end{table*}

\section{Results}
\label{sec:results}

\begin{figure}[t]
  \centering
  \includegraphics[width=0.9\columnwidth]{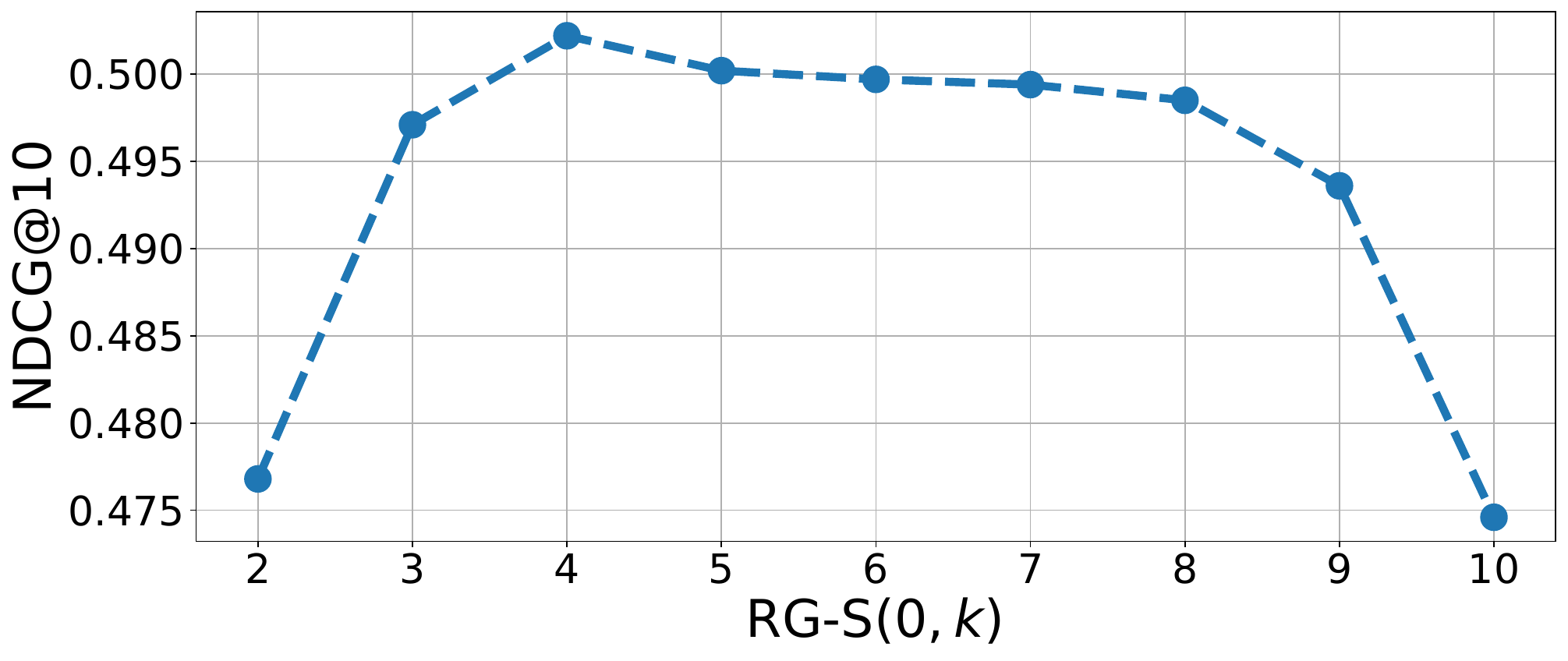}
  \caption{
  Average NDCG@10 on 8 BEIR data sets with different $k$ in rating scale $0$-to-$k$. 
  }
  \label{fig:rg_ks}
\end{figure}

\paragraph{Overall performance.}
Table~\ref{tab:overall_comparison} summarizes the overall comparison results.
It can be seen that prompting LLMs with fine-grained relevance labels achieves substantially higher performance than binary relevance labels (RG-YN, RG-2L).
For example, RG-3L on average achieves +2\% improvement in NDCG@10 compared with RG-2L and RG-YN.
RG-S$(0,4)$ which uses the rating scale 0 to 4 in the prompt also achieves similar improvement.
Note that even on data sets with binary ground-truth labels (e.g., SciFact), using fine-grained relevance labels still achieves substantial improvement.
This suggests that the improvement is not merely a result of matching the actual ground-truth relevance levels of the data set.

There are a few potential explanations for the observed improvement. 
One explanation is that the estimated relevance becomes more accurate as we aggregate more log-likelihood scores of multiple relevance labels.
Another is that the fine-grained relevance labels in the prompt help the LLMs to develop a more nuanced understanding of relevance.
We conduct more experiments to further explore these explanations.

\paragraph{Number of relevance labels.}
We first explore the effect of using different number of relevance labels.
Table~\ref{tab:overall_comparison} demonstrates that when using RG-$k$L, RG-4L performance is on par with RG-3L, suggesting that adding more relevance levels does not always improve the performance when using textual fine-grained relevance labels.

We also plot how the performance changes with regard to $k$ for the rating scale prompting method RG-S$(0,k)$ in Figure~\ref{fig:rg_ks}.
It shows that the performance from RG-S$(0,4)$ to RG-S$(0,8)$ remain similar.
This again suggests that using more fine-grained relevance labels does not further improve the performance.
Furthermore, performance declines for even larger $k$ such as RG-S$(0,9)$ and RG-S$(0,10)$.
This potentially indicates that LLMs struggle to understand prompts with excessive granularity~\cite{thawani-etal-2021-representing}.

Notably, the performance trend in Figure~\ref{fig:rg_ks} remains consistent across datasets regardless of varying granularity of ground-truth label (Appendix~\ref{sec:appendix_complete_results}).
This illustrates that, in practice, the performance gains are robust to a wide range of $k$ selections.

\begin{table}[t]
    \centering
    \caption{
        Comparing ranking score derivation strategies measured by average NDCG@10 on BEIR data sets. 
    }
    \scalebox{0.8}{
        \begin{tabular}{@{\hskip4pt}c|ccc@{\hskip4pt}}
        \toprule
        Prompts & Generated & Likelihood-ER & Likelihood-PR \\
        \midrule
         RG-3L & 0.3989 & 0.4992 & 0.5005 \\
         RG-4L & 0.4259 & 0.4969 & 0.4934 \\
         RG-S$(0,4)$ & 0.4445 & 0.5022  & 0.4988 \\
        \bottomrule
        \end{tabular}
    }
    \label{tab:ranking_score_strategy_comparison}
\end{table}

\paragraph{Ranking score derivation.}
We compare different strategies for deriving ranking scores.

Some existing work on LLM assessors~\cite{faggioli2023perspectives,thomas2023large} directly use the generated labels or scores without using the likelihood.
Technically, we can also rank documents directly based on the labels or scores parsed from the string outputs generated by LLMs.
We include this method in our comparison, denoted as ``Generated''.

Additionally, we compare the two strategies proposed in Section~\ref{sec:ranking_score_derivation}: expected relevance values (Likelihood-ER) and peak relevance likelihood (Likelihood-PR), both of which derive ranking scores from the predicted log-likelihood of LLMs.

The comparison results are presented in Table~\ref{tab:ranking_score_strategy_comparison}.
It is clear that directly using the generated labels or scores results in lower ranking performance compared to deriving scores from the log-likelihood, as it tends to introduce ties between documents.
On the other hand, peak relevance likelihood (Likelihood-PR) achieves very close performance to expected relevance values (Likelihood-ER) in most methods, despite only using the log-likelihood of one relevance label.
This suggests that the improvement brought by scoring fine-grained relevance labels cannot be simply explained by improved accuracy of estimated relevance by \textit{using more samples}.
Instead, it is possible that including fine-grained relevance labels within the prompt may signal LLMs to attend to the subtle relevance differences.

\begin{figure}[t]
  \centering
  \subfigure[RG-2L vs. RG-S$(0,4)$]{
    \label{subfig:rg2l_vs_rg4s}
    \includegraphics[width=0.47\columnwidth]{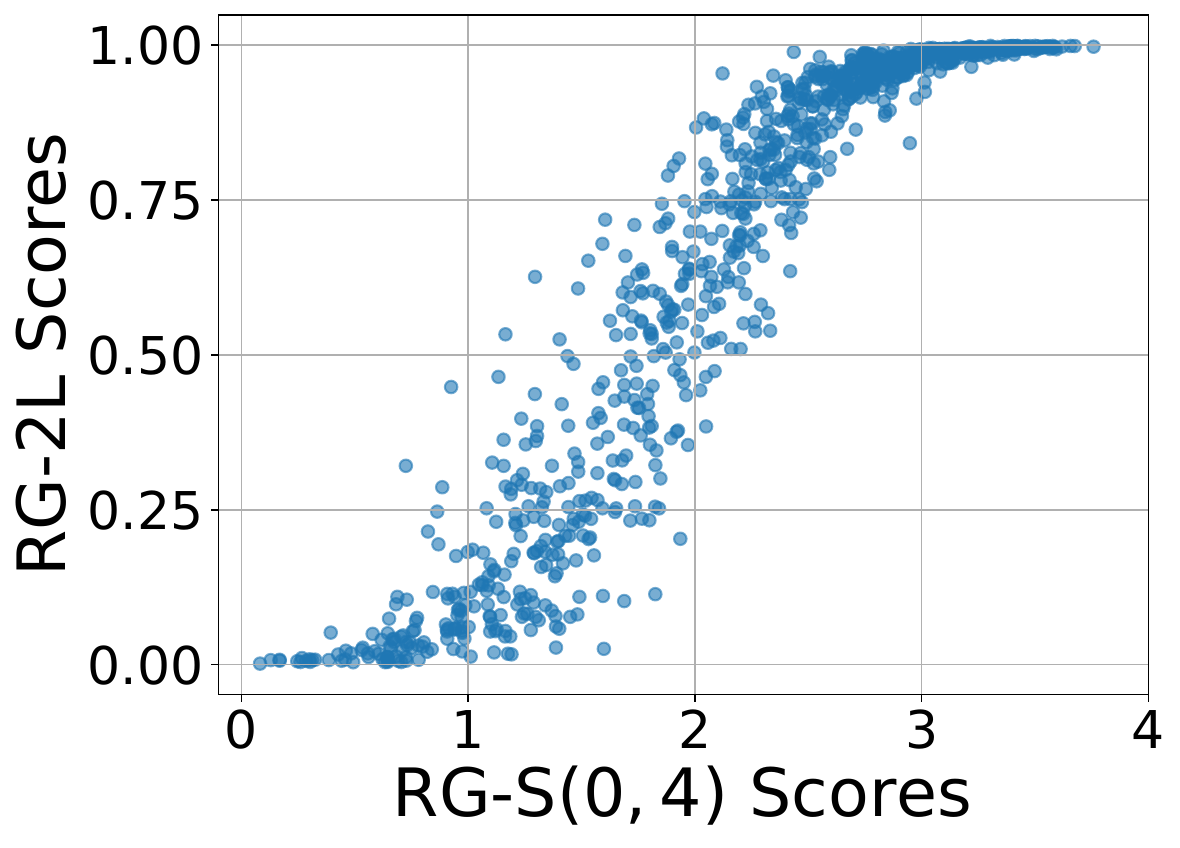}
  }
  \subfigure[RG-3L vs. RG-S$(0,4)$]{
    \label{subfig:rg3l_vs_rg4s}
    \includegraphics[width=0.46\columnwidth]{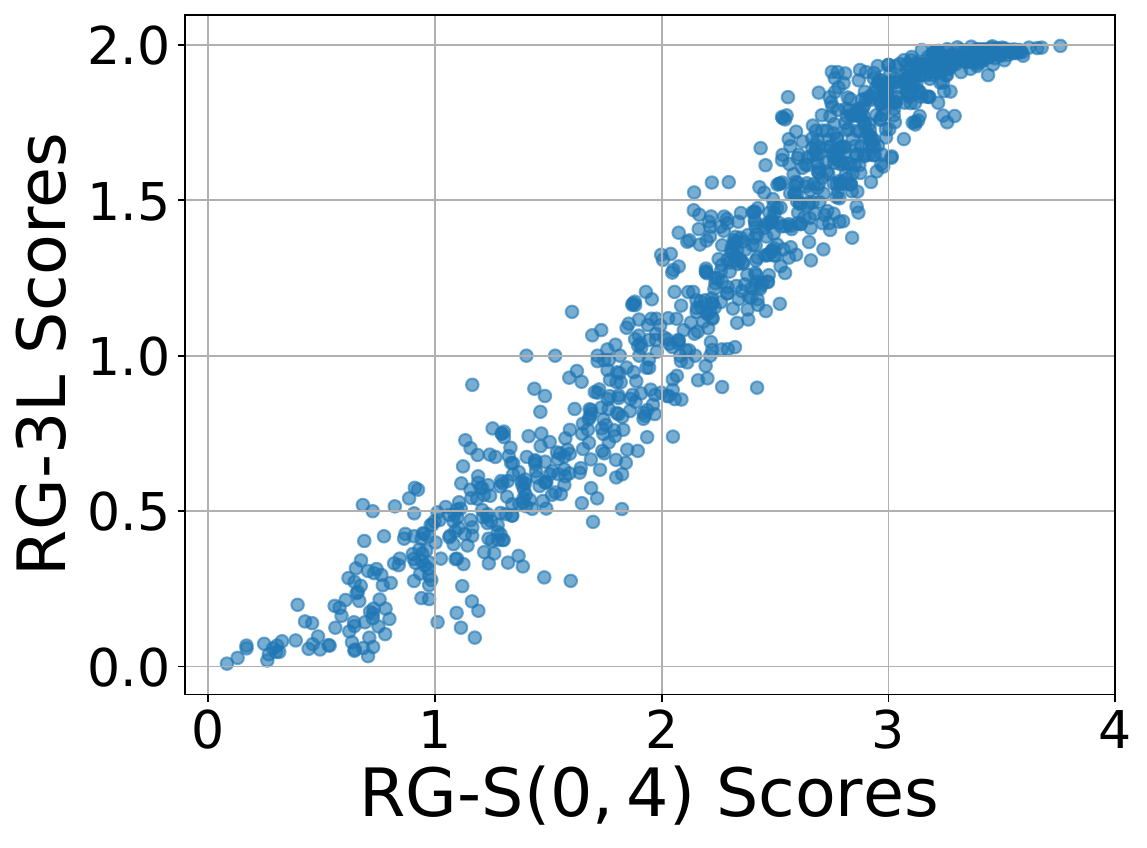}
  }
  \caption{\label{fig:size}
  Comparing ranking score distribution of different methods on the Covid data set.
  }
  \label{fig:score_scattered_plot}
\end{figure}

\paragraph{Score distribution comparison.}
We compare the score distributions of different methods to gain deeper insight into how fine-grained relevance labels influence performance.
Figure~\ref{fig:score_scattered_plot} presents a scatter plot of ranking scores (Likelihood-ER) from two methods for a random sample of query-document pairs in the Covid data set.

Figure~\ref{subfig:rg2l_vs_rg4s} demonstrates that RG-2L's ranking scores are mostly positively correlated with RG-S$(0,4)$'s (Figure~\ref{subfig:rg2l_vs_rg4s}), but struggles to distinguish query-document pairs with higher scores from RG-S$(0,4)$ and scores them almost equally with scores close to $1.0$.
This indicates that LLMs can differentiate better among higher-ranked relevant documents with fine-grained relevance labels.
In contrast, the ranking scores from RG-3L and RG-S$(0,4)$ (Figure~\ref{subfig:rg3l_vs_rg4s}) exhibit strong correlation almost throughout the entire range.
Correspondingly, RG-3L and RG-S$(0,4)$ also achieve similar ranking performance on this data set.

\section{Conclusion}
\label{sec:conclusion}

We explore pointwise zero-shot LLM rankers which score fine-grained relevance labels (e.g., ``Somewhat Relevant'') instead of binary labels. 
We propose to either provide intermediate relevance labels such as ``Somewhat Relevant'' as additional choices for the LLM or ask the LLM to rate the relevance between query-document pairs using a rating scale.
Then we aggregate the LLM likelihood scores of different relevance labels into ranking scores to rank the documents.
Further experiments illustrate that the performance gains are not solely attributable to more precise relevance estimation by using more samples,
as only using the log-likelihood of one relevance labels can also achieve similar performance gain. 
Instead, it is possible that the inclusion of fine-grained relevance labels in the prompt may guide LLMs to better differentiate documents, especially those ranked at the top. 

We believe that this approach can be extended beyond information retrieval to many other applications~\cite{liu2023g}.
For example, the same method can be applied for recommendation~\cite{fan2023recommender,wu2023survey}, where the LLM is asked to rate how likely a user would buy an item.

\section{Limitations}
\label{sec:limitations}

In this work, we assume that the predicted likelihood for any generated text can be accessed.
However, we are aware that this might not always be true for many proprietary LLMs where users can only call with specific APIs.

Our study is also limited to \emph{ranking} performance of LLMs, without further evaluation or analysis on whether our prompts can also improve LLM assessors.
Higher ranking performance does not always translate to higher relevance calibration performance~\cite{cohen2021not,faggioli2023perspectives,thomas2023large}, as the metrics have different emphasis.
It is possible that one needs to apply an appropriate transformation on the derived ranking scores from LLM likelihoods to achieve the best relevance calibration performance, which can be non-trivial.
We believe this is an intriguing research direction as it can further broaden the application~\cite{bahri2020choppy,shtok2012predicting} of the proposed methods.

\bibliography{references}
\newpage

\appendix

\section{Alternative Relevance Labels}

We replace the relevance labels with other phrases to examine how the performance changes.
For RG-2L, we replace ``Not Relevant'' with ``Irrelevant''; for RG-3L, we replace ``Somewhat Relevant'' with ``Partially Relevant''. 

The results are shown in Table~\ref{tab:alternative_relevance_levels}.
Regardless of using different textual representations of relevance labels, RG-3L consistently outperforms RG-2L. 
This suggests that the discovery in this paper is generalizable to different choices of textual relevance labels.
Another observation is that RG-2L performance varies slightly more than RG-3L performance.
This might indicate that RG-3L is more robust to different wording of relevance labels.

\begin{table}[h!]
    \centering
    \caption{
        Comparing ranking performance with different textual relevance labels. Measured by average NDCG@10 on BEIR data sets. 
    }
    \scalebox{0.9}{
        \begin{tabular}{c|p{0.6\columnwidth}|c}
        \toprule
         Method & Relevance Labels & Average \\
        \midrule
        \multirow{2}{*}{RG-2L} & ``Irrelevant'', ``Relevant'' &  0.4717 \\
        \cmidrule{2-3}
                               & ``Not Relevant'', ``Relevant'' & 0.4789 \\
        \midrule
        \multirow{4}{*}{RG-3L} & ``Not Relevant'', ``Partially Relevant'', ``Highly Relevant'' &  \multirow{2}{*}{0.4975} \\
        \cmidrule{2-3}
                               & ``Not Relevant'', ``Somewhat Relevant'', ``Highly Relevant'' & \multirow{2}{*}{0.4992} \\
        \bottomrule
        \end{tabular}
    }
    \label{tab:alternative_relevance_levels}
\end{table}

\begin{figure}[ht]
  \centering
  \includegraphics[width=0.9\columnwidth]{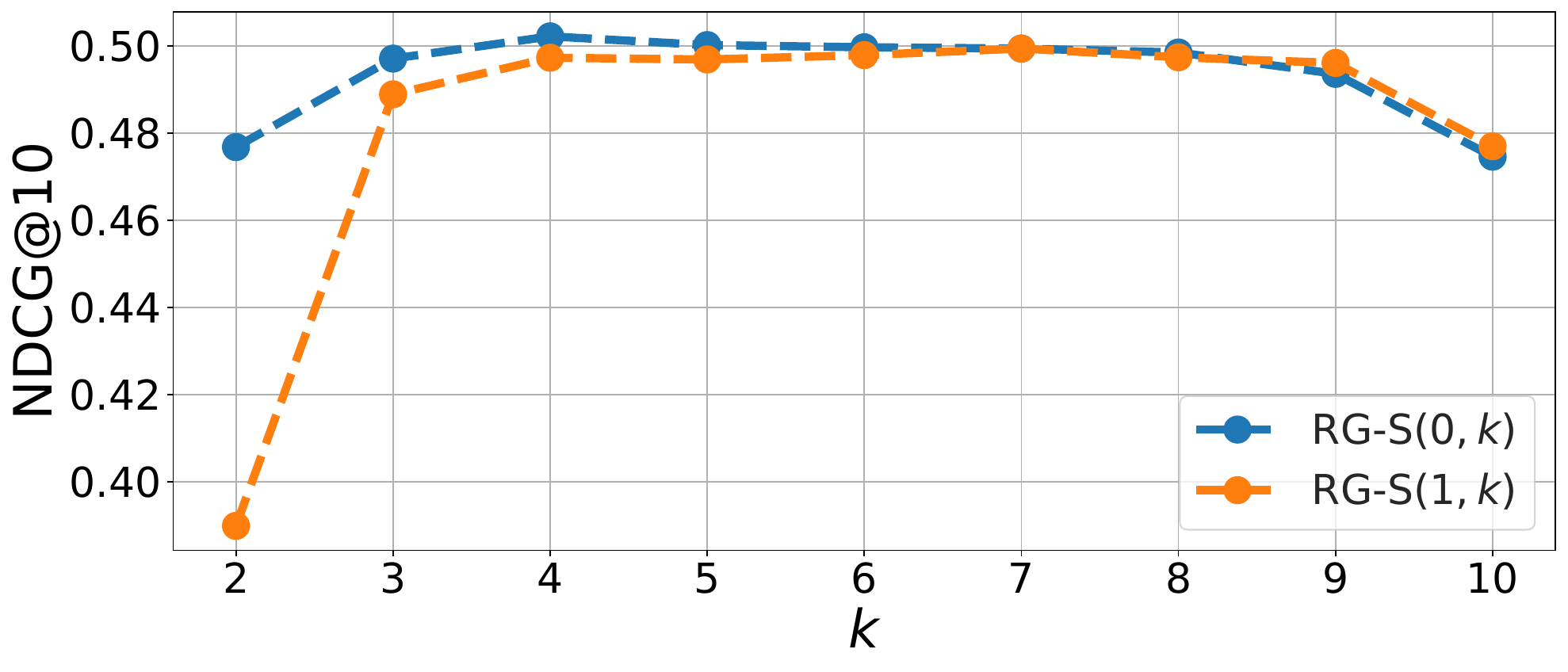}
  \caption{
  Comparing rating scale relevance generation with different prompts.
  }
  \label{fig:rg_ks_0tok_vs_1tok}
\end{figure}

We also experiment with different rating scale formulation.
Instead of prompting the LLM to rate the relevance from $0$ to $k$, we also try to ask the LLM to rate the relevance from $1$ to $k$, denoted as RG-S$(1,k)$. 
We plot the average NDCG@10 performance in Figure~\ref{fig:rg_ks_0tok_vs_1tok}.

The performance of both methods do not differ much when $k$ is larger than 4. 
But not providing the ``0'' option substantially hurt the performance when $k$ is lower than or equal to 3.
This might also suggest that using the rating scale from $0$ to $k$ is slightly more robust.

\section{In-Depth Score Distribution}
\begin{figure*}[t]
  \centering
  \includegraphics[width=1.85\columnwidth]{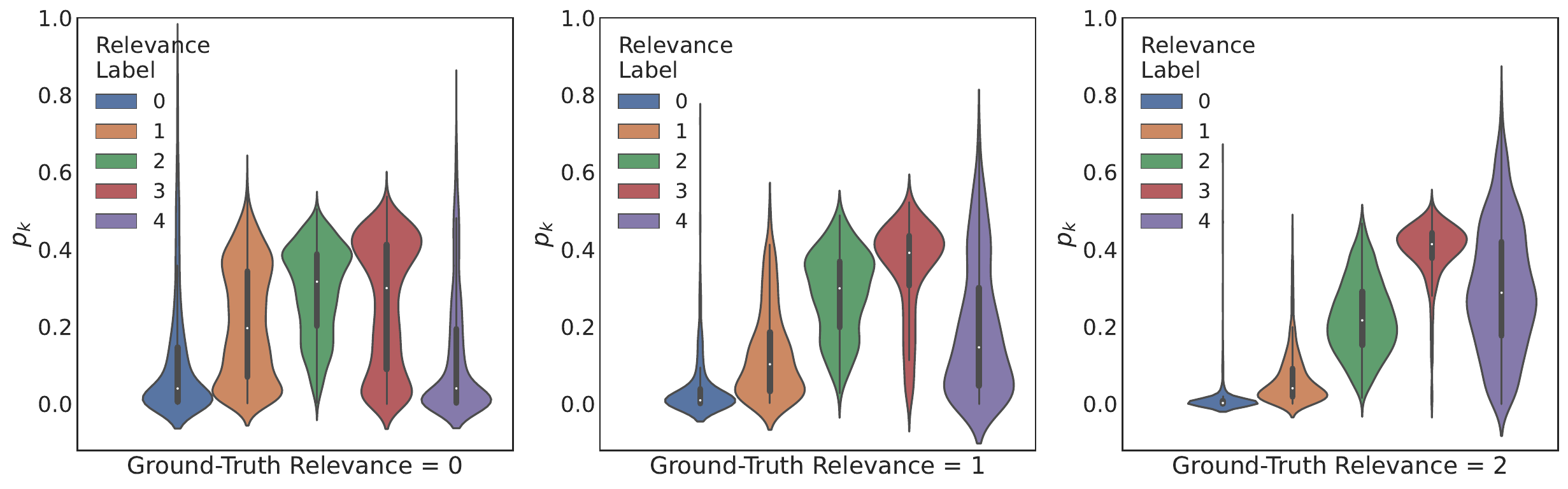}
  \caption{Distribution of marginal probability $p_k$ of each relevance label in RG-S$(0,4)$ for query-document pairs with different ground-truth labels on Covid data set}
  \label{fig:pk_distribution_rg_s04}
\end{figure*}

\begin{figure*}[t]
  \centering
  \includegraphics[width=1.85\columnwidth]{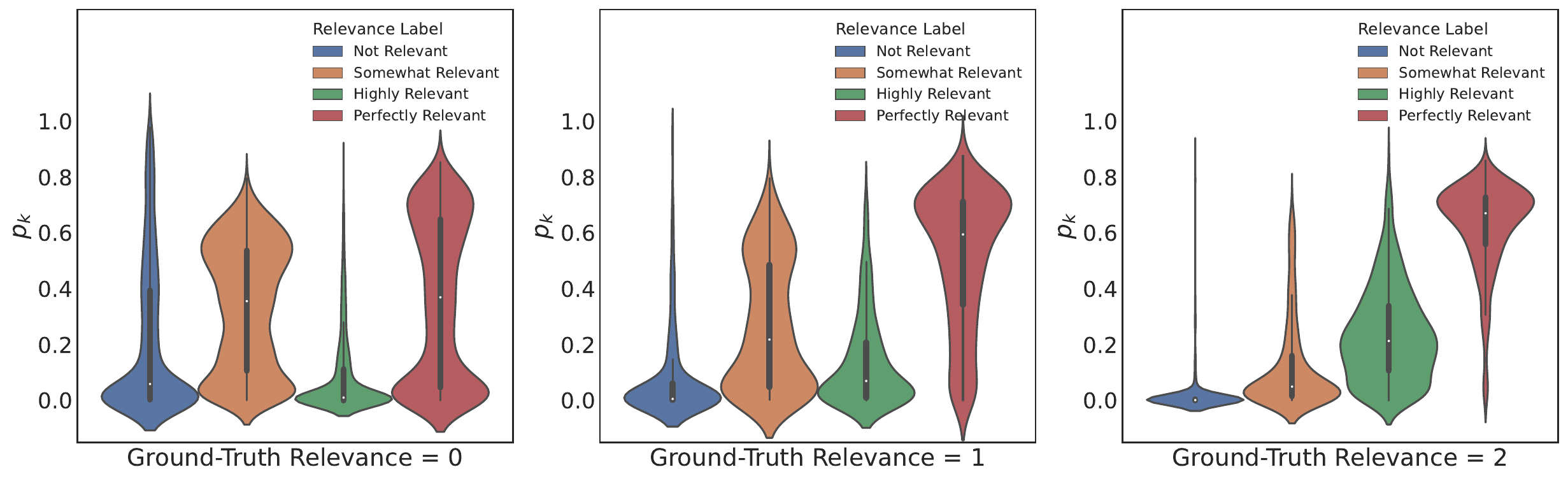}
  \caption{Distribution of marginal probability $p_k$ of each relevance label in RG-4L for query-document pairs with different ground-truth labels on Covid data set}
  \label{fig:pk_distribution_rg_4l}
\end{figure*}

We plot the in-depth score distribution of our methods.
Specifically, we group the query-document pairs in Covid data set by different ground-truth relevance.
We then denote $p_k$ as the random variable of the marginal probability $p_{i,k}$ derived for different query-document pairs $(q, d_i)$.
We the plot the estimated distribution of $p_k$ for each relevance label $l_k$ respectively.
Figure~\ref{fig:pk_distribution_rg_s04} and~\ref{fig:pk_distribution_rg_4l} shows the results on Covid data set when we use RG-S$(0,4)$ and RG-4L respectively.
The ground-truth relevance of Covid data set is 0, 1 or 2.

In Figure~\ref{fig:pk_distribution_rg_s04}, we observe that the distributions of marginal probability $p_k$ of relevance label ``0'', ``1'' and ``2'' shift down towards 0 as the ground-truth relevance increases. 
Meanwhile, the distributions of $p_k$ across relevance label ``3'' and ``4'' shift up towards 1. In Figure~\ref{fig:pk_distribution_rg_4l}, we found a similar trend where the distributions of marginal probability $p_k$ of ``Not Relevant'' and ``Somewhat Relevant'' shift down towards 0 as the ground-truth relevance increases, while the distributions of $p_k$ across ``Highly Relevant'' and ``Perfectly Relevant'' shift up towards 1. 
This reveals how our expected relevance values (ER) methods works in practice, and also given us hints on how peak relevance likelihood (PR) alone works based on the distribution shift of the peak relevance label.

\section{Varying Assigned Relevance Values}
We also investigate how the user provided relevance values $y_k$'s make a difference to the ranking
performance. 
We use RG-3L as the example. 
We fix $y_0=0$ for ``Not Relevant'' and $y_2=2$ for ``Highly Relevant'', but vary the relevance value $y_1$ for ``Somewhat Relevant'' between $y_0$ and $y_2$.
We evaluate the average NDCG@10 on the 8 BEIR data sets and presents the results in Table~\ref{tab:relevance_values}.

As $y_1$ varies, the average NDCG@10 does not change substantially when $y_1$ decreases.
Even when $y_1=y_0$, the NDCG@10 performance remains high. 
This is expected as NDCG@10 focuses on the top-ranked items, 
thus changing the relevance values of intermediate relevance labels may not significantly change the top-ranked items.

In contrast, when $y_1=y_2$, the performance drops significantly to about the same level as RG-2L.
This might indirectly explain why RG-2L performance is worse than RG-3L, as it might not be able to distinguish partially relevant and highly relevant documents.

\begin{table}[!t]
    \centering
    \caption{
        Comparing ranking performance with different relevance values $y_k$'s. Measured by average NDCG@10 on BEIR data sets. 
    }
    \scalebox{0.9}{
        \begin{tabular}{c|c|c}
        \toprule
         Method & $[y_0, y_1, y_2]$ & Average \\
        \midrule
        RG-3L & $[0.00, 0.00, 2.00]$ & 0.5000 \\
        RG-3L & $[0.00, 0.50, 2.00]$ & 0.5000 \\
        RG-3L & $[0.00, 1.00, 2.00]$ & 0.4992 \\
        RG-3L & $[0.00, 1.50, 2.00]$ & 0.4990 \\
        RG-3L & $[0.00, 2.00, 2.00]$ & 0.4779 \\
        \bottomrule
        \end{tabular}
    }
    \label{tab:relevance_values}
\end{table}

\section{Experiments on Other LLMs}
\label{sec:other_llms}

To verify the generalizability of our proposed method,  
we also conduct experiments on two other LLMs.
We use FLAN PaLM2 XS, which is a smaller alternative of FLAN PaLM2 S.
We also use FLAN UL2~\cite{tay2022ul2}, which is an open-sourced LLM with 20B parameters.
The results are presented in Table~\ref{tab:xs_comparison}.
We observe similar results where scoring fine-grained relevance labels (RG-3L, RG-S$(0,4)$) can achieve better average performance than scoring binary labels (RG-2L).
This shows that our method can generalize to different LLMs.

\begin{table*}[ht!]
    \centering
    \caption{
        Overall ranking performances of FLAN PaLM2 XS and FLAN UL2 measured by NDCG@10 on BEIR data sets. 
        The best performances are bolded.
    }
    \scalebox{0.75}{
        \begin{tabular}{c|c|cccccccc|c}
        \toprule
        Model & Method & Covid & Touche & DBPedia & SciFact & Signal & News & Robust04 & NFCorpus & Average \\
        \midrule
        \multirow{5}{*}{FLAN PaLM2 XS} 
        & RG-2L & 0.7769 & 0.2549 & 0.4228 & 0.6826 & 0.2892 & 0.4229 & \textbf{0.4947} & 0.3756 & 0.4649 \\
        & RG-3L & 0.7936 & 0.2554 & 0.4235 & 0.6810 & 0.2931 & \textbf{0.4374} & 0.4933 & \textbf{0.3777} &  0.4694 \\
        & RG-4L & 0.7969 & 0.2598 & 0.4277 & 0.6681 & 0.3004 & 0.4326 & 0.4772 & 0.3773 & 0.4675 \\
        \cmidrule{2-11}
        & RG-S$(0,2)$ & 0.7819 & 0.2535 & 0.4141 & \textbf{0.7135} & 0.2791 & 0.4356 & 0.4579 & 0.3711 &  0.4633 \\
        & RG-S$(0,4)$ & \textbf{0.8119} & \textbf{0.2885} & \textbf{0.4386} & 0.7102 & \textbf{0.3097} & 0.4341 & 0.4559 & 0.3763 &  \textbf{0.4781} \\
        \midrule
        \multirow{5}{*}{FLAN UL2} 
        & RG-2L & 0.7769 & \textbf{0.2737} & 0.4047 & 0.5626 & 0.2822 & 0.4573 & 0.5421 & 0.3756 & 0.4594 \\
        & RG-3L & 0.7998 & 0.2555 & 0.4303 & 0.7007 & 0.2928 & 0.4698 & \textbf{0.5582} & 0.3757 & 0.4853 \\
        & RG-4L & \textbf{0.8030} & 0.2477 & \textbf{0.4336} & 0.7186 & 0.3047 & \textbf{0.4710} & 0.5575 & \textbf{0.3775} & 0.4892 \\
        \cmidrule{2-11}
        & RG-S$(0,2)$ & 0.7915 & 0.2546 & 0.4252 & 0.7341 & 0.2997 & 0.4700 & 0.5497 & 0.3702 & 0.4869  \\
        & RG-S$(0,4)$ & 0.7969 & 0.2641 & 0.4325 & \textbf{0.7391} & \textbf{0.3129} & 0.4557 & 0.5454 & 0.3708 & \textbf{0.4897}  \\
        \bottomrule
        \end{tabular}
    }
    \label{tab:xs_comparison}
\end{table*}

\begin{table*}[!t]
    \centering
    \caption{
        Overall ranking performances measured by NDCG@10 on BEIR data sets. 
    }
    \scalebox{0.75}{
        \begin{tabular}{cc|cccccccc|c}
        \toprule
         Method & Model &  Covid & Touche & DBPedia & SciFact & Signal & News   & Robust04 & NFCorpus & Average \\
        \midrule
         BM25   & N/A   & 0.5947  & 0.4422  & 0.3180 & 0.6789 & 0.3305 & 0.3952 & 0.4070    & 0.3075 & 0.4342 \\
        \midrule
        QG & FLAN PaLM2 S &  0.7357	& 0.2408& 0.3773& 0.7495& 0.2872& 0.4156& 0.4651	&0.3673& 0.4548 \\
         RG-YN & FLAN PaLM2 S& 0.7897 & 0.2427 & 0.3696 & 0.6958 & 0.3196 & 0.4588 & 0.5656 & 0.3743 & 0.4770 \\
        \midrule
         RG-2L-ER & FLAN PaLM2 S& 0.7949 & 0.2411 & 0.3590 & 0.7290 & 0.2996 & 0.4623 & 0.5636 & 0.3814 & 0.4789 \\
         RG-3L-ER & FLAN PaLM2 S& 0.8065 & 0.2650 & 0.4013 & 0.7671 & 0.3142 & 0.4890 & 0.5660 & 0.3849 & 0.4992 \\
         RG-4L-ER & FLAN PaLM2 S& 0.8063 & 0.2388 & 0.4033 & 0.7766 & 0.3184 & 0.4884 & 0.5635 & 0.3801 & 0.4969 \\
        \midrule
         RG-2L-PR & FLAN PaLM2 S & 0.7874 & 0.2482 & 0.3435 & 0.7230 & 0.2819 & 0.4619 & 0.5647 & 0.3706 & 0.4726 \\
         RG-3L-PR & FLAN PaLM2 S & 0.8065 & 0.2634 & 0.4032 & 0.7745 & 0.3202 & 0.4816 & 0.5681 & 0.3860 & 0.5005 \\
         RG-4L-PR & FLAN PaLM2 S & 0.8076 & 0.2354 & 0.4050 & 0.7772 & 0.3121 & 0.4712 & 0.5561 & 0.3824 & 0.4934 \\
        \midrule
         RG-S$(0,2)$-ER & FLAN PaLM2 S & 0.7760 & 0.2695 & 0.3709 & 0.6921 & 0.3034 & 0.4677 & 0.5557 & 0.3787 & 0.4768 \\
         RG-S$(0,3)$-ER & FLAN PaLM2 S & 0.7936 & 0.2720 & 0.4092 & 0.7434 & 0.3240 & 0.4817 & 0.5662 & 0.3868 & 0.4971 \\
         RG-S$(0,4)$-ER & FLAN PaLM2 S & 0.8048 & 0.2757 & 0.4190 & 0.7521 & 0.3301 & 0.4790 & 0.5668 & 0.3901 & 0.5022 \\
         RG-S$(0,5)$-ER & FLAN PaLM2 S & 0.8088 & 0.2702 & 0.4217 & 0.7475 & 0.3266 & 0.4734 & 0.5666 & 0.3871 & 0.5002 \\
         RG-S$(0,6)$-ER & FLAN PaLM2 S & 0.7898 & 0.2720 & 0.4260 & 0.7529 & 0.3288 & 0.4734 & 0.5687 & 0.3864 & 0.4997 \\ 
         RG-S$(0,7)$-ER & FLAN PaLM2 S & 0.7873 & 0.2695 & 0.4225 & 0.7557 & 0.3263 & 0.4848 & 0.5659 & 0.3831 & 0.4994 \\ 
         RG-S$(0,8)$-ER & FLAN PaLM2 S & 0.7971 & 0.2730 & 0.4254 & 0.7463 & 0.3239 & 0.4722 & 0.5647 & 0.3853 & 0.4985 \\ 
         RG-S$(0,9)$-ER & FLAN PaLM2 S & 0.7910 & 0.2746 & 0.4160 & 0.7465 & 0.3017 & 0.4679 & 0.5644 & 0.3871 & 0.4936 \\
         RG-S$(0,10)$-ER & FLAN PaLM2 S & 0.7576 & 0.2496 & 0.3738 & 0.7310 & 0.2771 & 0.4779 & 0.5642 & 0.3655 & 0.4746 \\
         \midrule
         RG-S$(0,2)$-PR & FLAN PaLM2 S & 0.7821 & 0.2735 & 0.3469 & 0.6954 & 0.2597 & 0.4540 & 0.5409 & 0.3752 & 0.4659 \\
         RG-S$(0,4)$-PR & FLAN PaLM2 S & 0.8036 & 0.2785 & 0.4221 & 0.7625 & 0.3168 & 0.4623 & 0.5559 & 0.3886 & 0.4988 \\
        \midrule
         monoT5 & Fine-tuned T5 XL & 0.8071 & 0.3241 & 0.4445 & 0.7657 & 0.3255 & 0.4849 & 0.5671 & 0.3897 & 0.5136 \\
         RankT5 & Fine-tuned T5 XL & 0.8200 & 0.3762 & 0.4419 & 0.7686 & 0.3180 & 0.4815 & 0.5276 & 0.3860 & 0.5150 \\
        \midrule
         RankGPT & GPT-3.5 Turbo & 0.7667  & 0.3618 &0.4447 &0.7043& 0.3212 & 0.4885 & 0.5062 & 0.3562 & 0.4937 \\
         PRP & UL2 & 0.7945 & 0.3789& 0.4647 & 0.7333 & 0.3520 & 0.4911 & 0.5343 & N/A & N/A \\
        \bottomrule
        \end{tabular}
    }
    \label{tab:all_comparison}
\end{table*}

\section{More Comparison Results}
\label{sec:appendix_complete_results}
We also include a more thorough comparison with other methods including:
\begin{itemize}
    \item BM25. The base retriever performance.
    \item monoT5~\cite{nogueira2020T5ranking}. A T5 XL model fine-tuned on MS MARCO data set for text ranking task and applied directly on the BEIR data sets.
    \item RankT5~\cite{zhuang2023rankt5}. An encoder-only model initialized with T5 XL but fine-tuned on MS MARCO data set using listwise softmax cross-entropy ranking loss and applied directly on the BEIR data sets.
    \item Pairwise Ranking Prompts (PRP)~\cite{qin2023prp}. A zero-shot pairwise LLM ranker which takes a query and two documents as input, and outputs which one is more relevant to the query. We include the best results of PRP which uses UL2 as the LLM and a sliding window strategy.
    \item RankGPT~\cite{sun2023chatgpt}. A zero-shot listwise LLM ranker which takes a query and a list of documents as input, and outputs an ordered list of documents based on their relevance. The method is used jointly with a sliding window strategy. We do not include the GPT-4 reranking number as it involves a second-stage ranking. 
\end{itemize}
We also include the detailed results of our proposed methods with different $k$ values, and different strategies to derive ranking scores.
Table~\ref{tab:all_comparison} illustrates the results. 

It is not surprising that our methods perform slightly worse than monoT5 or RankT5 as they are fine-tuned for the text ranking task on MS MARCO data set. 
However, it is encouraging to see our prompting method substantially shrinks the gap between zero-shot LLM rankers and RankT5. 

Our methods can also perform slightly better than the single-stage RankGPT.
However, note that the LLM used in these experiments are different, so the difference might also be explained by the model difference.

\begin{figure}[t]
  \centering
  \includegraphics[width=0.9\columnwidth]{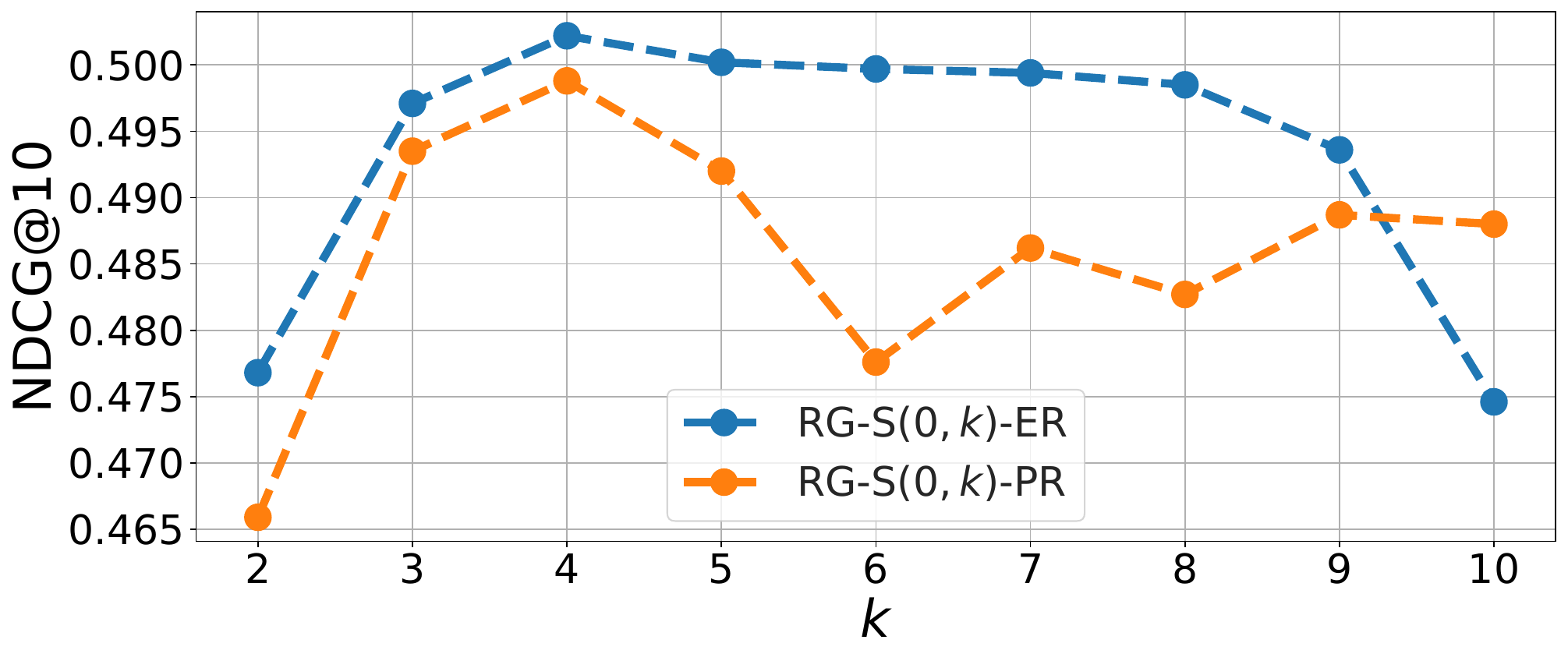}
  \caption{
  Comparing rating scale relevance generation with different strategies to derive ranking scores.
  }
  \label{fig:rg_ks_er_vs_pr}
\end{figure}

Figure~\ref{fig:rg_ks_er_vs_pr} also plots the performance of rating scale methods ranking score derivation methods. 
It can be observed that the ranking performance of using PR to derive ranking scores is more sensitive to the selection of $k$ than using ER.

\begin{table}[!t]
    \centering
    \caption{
        Comparing ranking performance instruction and in-context learning. Measured by average NDCG@10 on BEIR data sets. 
    }
    \scalebox{0.9}{
        \begin{tabular}{l|c}
        \toprule
         Method & Average \\
        \midrule
         RG-2L & 0.4789 \\
         + Instructions & 0.4914 \\
         + Instructions + 4-shot ICL & 0.4914 \\
        \midrule
         RG-3L & 0.4992 \\
         + Instructions & 0.5034 \\
         + Instructions + 4-shot ICL & 0.5046 \\
        \bottomrule
        \end{tabular}
    }
    \label{tab:instruction_and_icl}
\end{table}

\section{Instructions and In-Context Learning}
We also try adding instructions and few-shot exemplars into the prompt.
For instructions, we directly add the definition of the relevance labels into the prompt.
The relevance label definitions are directly copied from TREC-DL 2020~\cite{trecdl2020}.
For RG-2L instructions we use the ``Irrelevant'' and ``Relevant'' labels;
for RG-3L instructions we use the ``Irrelevant'', ``Relevant'' and ``Highly Relevant'' labels.
We also change the relevance labels accordingly to align with the instructions.

In addition to instructions, we also try to include few-shot exemplars to leverage the model's in-context learning capabilities.
We include 4-shot exemplars, which are randomly sampled from TREC-DL 2020 data sets.
We sampled 2 ``Irrelevant'', 1 ``Relevant'' and 1 ``Perfectly Relevant'' query-document pairs.
To align with the instructions, for RG-2L we label both ``Relevant'' and ``Perfectly Relevant'' exemplar query-document pairs as ``Relevant''; for RG-3L we label the ``Perfectly Relevant'' pair as ``Highly Relevant''.

The results are shown in Table~\ref{tab:instruction_and_icl}.
Adding instructions improves both RG-2L and RG-3L, while RG-3L still remains +1.2\% better than RG-2L.
Further adding exemplars on top of the instructions does not improve much, possibly due to the distribution discrepancy between TREC-DL and BEIR.

\onecolumn

\section{Prompts}
\label{sec:appendix_prompts}
In this section, we provide the prompts we used for each method:

\subsection{Query Generation (QG)}

We use the following prompt for our QG experiments. 
We find this prompt performs better empirically for zero-shot QG LLM rankers than the prompt used in existing works~\cite{sachan2022improving}.
\begin{tcolorbox}
I will check whether what you said could answer my question. \\

You said: \code{\{document\}}

I googled: \code{\{query\}} 

\end{tcolorbox}

\subsection{Binary Relevance Generation (RG-YN)}
We use the following prompt for our RG-YN experiments. 
We find this prompt performs better empirically than the prompt used originally by~\citet{liang2022holistic}, ~\citet{sun2023chatgpt} and~\citet{qin2023prp}. 

\begin{tcolorbox}
For the following query and document, judge whether they are relevant. Output ``Yes'' or ``No''. \\

Query: \code{\{query\}}

Document: \code{\{document\}}

Output:
\end{tcolorbox}

\subsection{2-Level Relevance Generation (RG-2L)}
\begin{tcolorbox}
For the following query and document, judge whether they are ``Relevant'', or ``Not Relevant''. \\

Query: \code{\{query\}}

Document: \code{\{document\}}

Output:
\end{tcolorbox}

\subsection{3-Level Relevance Generation (RG-3L)}
\begin{tcolorbox}
For the following query and document, judge whether they are ``Highly Relevant'', ``Somewhat Relevant'', or ``Not Relevant''. \\

Query: \code{\{query\}}

Document: \code{\{document\}}

Output:
\end{tcolorbox}

\subsection{4-Level Relevance Generation (RG-4L)}
\begin{tcolorbox}
For the following query and document, judge whether they are ``Perfectly Relevant'', ``Highly Relevant'', ``Somewhat Relevant'', or ``Not Relevant''. \\

Query: \code{\{query\}}

Document: \code{\{document\}}

Output:
\end{tcolorbox}

\subsection{Rating Scale Relevance Generation (RG-S$(0, k)$)}
\begin{tcolorbox}
From a scale of 0 to \code{\{k\}}, judge the relevance between the query and the document. \\

Query: \code{\{query\}}

Document: \code{\{document\}}

Output:
\end{tcolorbox}
\end{document}